\documentclass[letterpaper,11pt]{article}

\usepackage{jheppub1} 

\usepackage[T1]{fontenc} 
\usepackage{soul}
\usepackage{amssymb}
\usepackage{amsmath}
\usepackage{amsfonts}
\usepackage{graphicx}
\usepackage{subfigure}
\usepackage{color}
\usepackage{dcolumn}
\usepackage{hyperref}
\numberwithin{equation}{section}

\newcommand{\be}{\begin{equation}}
\newcommand{\ee}{\end{equation}}
\newcommand{\bea}{\begin{eqnarray}}
\newcommand{\eea}{\end{eqnarray}}
\newcommand{\nn}{\nonumber}
\newcommand{\rd}{\partial}

 \def\cB{{\cal B}} 
  \def\cF{{\cal F}}
 \def\cH{{\cal H}} 
  
 \def\cN{{\cal N}} \def\cO{{\cal O}}

 \def\cZ{{\cal Z}}

\def\a{\alpha}      
\def\b{\beta}       
\def\g{\gamma}    
\def\d{\delta}  \def\D{\Delta}

\def\k{\kappa}

\def\o{\omega}

\def\s{\sigma}  
\def\t{\tau}

\newcommand{\vev}[1]{{\left< {#1} \right>}}

\newcommand{\prt}[1]{{\left( {#1} \right)}}
\newcommand{\eq}[1]{(\ref{#1})}
\newcommand{\la}[1]{\label{#1}}

\makeatletter
\@addtoreset{equation}{section}
\makeatother

\newcommand{\comment}[1]{}

\title{\boldmath  Heavy Quark Diffusion in Strongly Coupled Anisotropic Plasmas}

\author{Dimitrios Giataganas${^1}$,}
\author{Hesam Soltanpanahi${^{2,3}}$}

\affiliation{${^1}$Department of Physics, University of Athens, 15771 Athens, Greece\\
${^2}$Institute of Physics, Jagiellonian University, Reymonta 4, 30-059 Krak\'{o}w, Poland\\
${^3}$National Institute for Theoretical Physics, School of Physics and Centre for Theoretical Physics,\\
University of the Witwatersrand, Wits, 2050, South Africa}

\emailAdd{dimitrios.giataganas@gmail.com}
\emailAdd{hesam@th.if.uj.edu.pl}

\abstract{
We study  the Langevin diffusion of a relativistic heavy quark in anisotropic strongly coupled theories in the local limit.  Firstly, we use the axion space-dependent deformed anisotropic $\cN=4$ sYM, where the geometry anisotropy is always prolate, while the pressure anisotropy may be prolate or oblate. For motion along the anisotropic direction we find that the effective temperature for the quark can be larger than the heat bath temperature, in contrast to what happens in the isotropic theory. The longitudinal and transverse Langevin diffusion coefficients depend strongly on the anisotropy, the direction of motion and the transverse direction considered. We analyze the anisotropy effects to the coefficients and  compare them to each other and to them of the isotropic theory.  

To examine the dependence of the coefficients on the type of the geometry,
we consider another bottom-up anisotropic model. Changing the geometry from prolate to oblate, certain diffusion coefficients interchange their behaviors.

In both anisotropic backgrounds we find cases that the transverse diffusion coefficient is larger than the longitudinal, but we find no negative excess noise.}

\preprint{WITS-CTP-125}

\begin{document}
\maketitle
\flushbottom

\section{Introduction}
The dynamics of the heavy quarks provide important information in the study of the Quark Gluon Plasma (QGP) created in the Heavy Ion Colliders. The relevant findings suggest that the QGP is strongly coupled \cite{Adams:2005dq,Adcox:2004mh,Shuryak:2004cy} and therefore a promising approach to study these phenomena is by the use of gauge/gravity correspondence \cite{adscft1,adscft2}, where a recent review may be found in \cite{adscftappl}.

The QGP goes through different phases in a short period of time. Before it reaches the isotropic phase, it goes through an anisotropic, both in momentum and phase space. The time period that the anisotropic phase lasts, is not yet specified accurately, and isotropization and thermalization is currently under intensive studies. Short times of order $2 fm$ are predicted using conformal viscus hydrodynamics where the values depend strongly on the initial conditions. However, holographic models predict lower times $\sim 0.3 fm$ \cite{Chesler:2010bi}. The anisotropic phase of the plasma is followed by a longer lasting isotropic phase. Several observables in this phase have been studied extensively using the gauge/gravity correspondence. Recently, these studies were extended in the anisotropic phase of the plasma \cite{aniso2, Mateos:2011ix, aniso1, Rebhan:2011vd,giataganasan, Chernicoff:2012iq,  Chernicoff:2012gu, Rebhan:2012bw, Chernicoff:2012bu,Fadafan:2012qu, Gynther:2012mw, Muller:2012uu, Patino:2012py, Chakraborty:2012dt, Fadafan:2013bva, Ali-Akbari:2013txa, Chakrabortty:2013kra, Jahnke:2013rca}, {}\footnote{Anisotropic hydrodynamical models are also attracting increasing attention eg.\cite{Ryblewski:2013jsa, Gahramanov:2012wz, Martinez:2012tu, Ryblewski:2012rr, Florkowski:2013lya}.} where a recent review may be found in \cite{Giataganas:2013lga}. Here, we extend them further by examining the Langevin dynamics of a moving quark in the anisotropic plasma.

The out of equilibrium heavy quarks go under a Brownian like motion with a stohastic force $\xi\prt{t}$, and provide observables that are important for the plasma, where a summary related to their physics can be found in \cite{Rapp:2009my}. Moreover a more mathematical approach to the Langevin diffusion coefficients is reviewed in \cite{Dunkel20091}. It has been an extensive study of the Langevin diffusion coefficients of heavy quarks in several gauge/gravity dualities initiated in \cite{Herzog:2006gh,Gubser:2006bz,Gubser:2006nz,CasalderreySolana:2006rq, CasalderreySolana:2007qw,deBoer:2008gu,Son:2009vu,Giecold:2009cg}, and further extended in \cite{Gursoy:2009kk,HoyosBadajoz:2009pv,Gursoy:2010aa,Kiritsis:2013iba}. A completely generic approach for a large class of theories using the membrane paradigm, was given very recently in \cite{Giataganas:2013hwa}.

In this paper we extend the study of the relativistic Langevin coefficients using the holography in the context of plasmas that are anisotropic. Our main purpose is to study the diffusion coefficients on the anisotropic theories and understand the possible physical implications of our results in the dual plasma. We are also motivated by the fact that in the anisotropic plasmas the universal inequality between the longitudinal and the transverse Langevin coefficients has been found to be violated \cite{Giataganas:2013hwa}, and we would like to understand better the conditions of this violation. Another motivation for our paper is that in \cite{Giataganas:2013hwa} was argued that the only possible way to obtain negative noise coefficients, is for the motion of the quark in anisotropic plasmas and we would like to examine here this possibility for different anisotropic models.

For our analysis we consider a fundamental string which has an endpoint at the UV boundary of the anisotropic backgrounds, representing the heavy moving quark. The string end point moves with a constant velocity $v$ equal to that of the heavy quark. The momentum flowing from the boundary to the bulk can be found and subsequently the force of the drag applied to the quark during its motion to the plasma is obtained. The direction of motion of the quark, affects the results of the magnitude of the drag force since the plasma is anisotropic. The further details of the calculations and the findings depend on the details of the anisotropic theory we study.
In our paper we consider two models to study the anisotropic motion, the top down space dependent axion deformed $\cN=4$ sYM \cite{aniso1} and the
bottom-up anisotropic model \cite{aniso2}. Both geometries contain one anisotropic space direction and an $SO\prt{2}$ isotropic plane. We  extensively analyze the coefficients in the top-down model in the whole range of anisotropies. A reason that bottom-up model is considered, is that it has the advantage to include prolate and oblate geometries, which is helpful to obtain a connection between the different types of the background geometry and certain Langevin coefficients.

The equations of motion for the trailing string for motion in both transverse and anisotropic directions have been studied in \cite{giataganasan,Chernicoff:2012iq}. Along each direction we find a different string solution stretching inside the bulk and that the 2-dim induced metric has a worldsheet black hole and a horizon at a radial point $u_0$. Therefore a worldsheet Hawking temperature $T_{ws}$ is associated to this black hole, which is in principle different to the heat bath temperature and it approaches it only when the quark moves non-relativistically. In the anisotropic case the exact position of the world-sheet horizon depends on the direction of motion, and therefore the corresponding world-sheet temperature as well. In the usual conformal case the $T_{ws}$ is less than the heat bath temperature in several setups, leading to holographic refrigerator systems \cite{Gursoy:2010aa,Nakamura:2013yqa}. In the anisotropic case we find that this inequality may be inverted, depending on the direction of motion of the quark, the degree of anisotropy and the speed of motion.

By considering the fluctuations of the trailing string we relate the Langevin coefficients to the thermal correlators. It turns out, as in the isotropic case, that these are thermal with the temperature $T_{ws}$ which was argued to correspond to the temperature measured by the quark moving in the plasma via the fluctuation-dissipation theorem \cite{Cugliandolo:1997nr,Gursoy:2010aa}. Then we study the spectral densities for low frequencies compared to the temperature, using the membrane paradigm. We review and apply the completely generic formalism developed in \cite{Giataganas:2013hwa} and additionally derive a generalization of the Einstein relation for generic theories. In the space dependent axion anisotropic model we study the parallel $\k_L$ and transverse $\k_T$ Langevin coefficients to the quark motion along the anisotropic direction and the transverse plane, for large and small anisotropies. We find their analytical relations in the later case. In the whole range of anisotropies we compare them each other and to the isotropic coefficients. We also analyze the cases where the universal inequality $\k_L\geq\k_T$ does not hold.  By considering the bottom-up model we take advantage of the fact that includes prolate and oblate geometries, depending on the background parameters, and observe the relation between the drag forces and the Langevin coefficients when going from oblate to prolate geometries. In particular by changing the geometry from oblate to prolate and vice versa certain Langevin coefficients for motion along the transverse and longitudinal directions interchange qualitative behaviors. Therefore we find a correlation of the type of the geometry and the Langevin coefficients, which does not necessary carry on to the pressure type anisotropy as the findings of the top-down model indicate.

Finally, we examine the possibility  of the negative excess noise in our anisotropic models and we find that in both models there is strictly positive excess noise. The conditions for negative excess  noise \cite{Giataganas:2013hwa} turn out to be very strict to get satisfied even for anisotropic theories.

The outline of the paper is as follows. In section \ref{section2} we review the completely generic formalism developed in \cite{Giataganas:2013hwa} for a string moving in a non-confining background. We also derive a generalization of the Einstein relation for generic theories.  In section \ref{section4} we introduce the anisotropic theories that we plan to study. Then in section \ref{section5} we study the Langevin coefficients in small and large anisotropies of the space dependent axion deformed anisotropic theory. In section \ref{section6}, we do the same in the bottom-up anisotropic model. In section \ref{section7} we comment on common results between these two models. Finally we conclude by discussing the implications of our results in the anisotropic theories, the violation of the universality relations and the absence of negative excess noise in section \ref{section8}. For presentation purposes some analytical results  for both models are given in the Appendices \ref{app1} and \ref{app:j}.

\section{Generic Study of the Trailing String and Setup}\la{section2}

We briefly review some of the generic results of \cite{Giataganas:2013hwa} focusing on the ones we need to apply to study the anisotropic theories.
We consider a background of the form
\be\la{meta1}
ds^2=G_{00}dt^2+G_{uu}du^2+G_{ii}dx_i^2~,
\ee
which is diagonal and allows the study of anisotropic cases. The metric components are functions of radial coordinate $u$, the boundary of the space is taken at $u\rightarrow 0$ and the element $G_{00}$ depends on the black hole horizon.
The trailing string corresponding to a quark moving on the boundary along
the chosen direction $x_p$, $p=1, 2, 3,$  with a constant velocity has the usual parametrization
\be
t=\t,~~~u=\sigma,~~~x_p=v~t+\xi(u),
\ee
and localized in the rest of dimensions.
Taking the Nambu-Goto action
\be
S_{NG}=-\frac{1}{2\,\pi\,\alpha'}\int\,d^2\sigma\,\sqrt{-g}~,
\ee
with $g_{\alpha \beta}$ being the induced world-sheet metric, we solve for  $\xi'$ in terms of the momentum flowing from the boundary to the bulk, which is a constant of motion
\be
\xi'^2=-G_{uu} C^2\,\frac{G_{00}+G_{pp}\,v^2} {G_{00}G_{pp}\prt{C^{2}+G_{00}G_{pp}}}~,\quad C:=2~\pi\,\alpha'\,\Pi^p_u~.
\ee
There is a critical point at which both numerator and denominator change their sign. This point $u_0$ is found by solving the equation
\be
G_{00}(u_0)=-G_{pp}(u_0)\,v^2~,\label{critical-point1}
\ee
where we have assumed $G_{uu}(u_0)\neq 0$. The corresponding drag force is
calculated at this point
\be
F_{drag,x_p}=-\frac{1}{2\pi\alpha'}\frac{\sqrt{-G_{00}(u_0)\,G_{pp}(u_0)}} {2\pi}=-\frac{v\, G_{pp}(u_0)}{2\pi\alpha'}~,
\ee
while the friction coefficient is defined by
\bea
F_{drag}=\frac{dp}{dt}=-\eta_D p,\qquad \eta_D=\frac{G_{pp}(u_0)}{2\pi\alpha' M_Q \gamma}~,\label{etaD}
\eea
where $p=M_Q\,v\gamma~,$ $\gamma:=\left(1-v^2\right)^{-1/2}$
and $M_Q$ is the mass of the heavy probe quark.

The world-sheet of the string has a horizon obtained by $g_{\tau\tau}(\s_h)=0$ and turns out to be the same with critical point $u_0$ .  They are obtained by solving the equation \eq{critical-point1}.
In order to find the effective temperature of the world-sheet horizon we diagonalize the world-sheet metric by as $d\t\rightarrow d\tilde{\t}= d\t-g_{\t\s}/g_{\t\t}~d\s$.
The diagonal metric components read 
\bea\la{m2}
h_{\tilde{\t}\tilde{\t}}=G_{00}+v^2\,G_{pp}~,\qquad
h_{\s\s}=
\frac{G_{00}G_{uu}G_{pp}}{G_{00}G_{pp}+C^2} ~.
\eea
The temperature then can be obtained following the usual process and is given by
\bea\label{wst1}
T_{ws}^2=\frac{1}{16\pi^2}
\left|\frac{G_{00}'^2-v^4\,G_{pp}'^2} {G_{00}G_{uu}}\right|\bigg|_{u=u_0}=\frac{1}{16\pi^2}
\Bigg|\frac{1}{G_{00} G_{uu}}\prt{G_{00}\,G_{pp}}' \prt{\frac{G_{00}}{G_{pp}}}'\Bigg|\bigg|_{u=u_0}~,
\eea
where in the first relation the velocity enters explicitly and the
second equality is written such that only the background
metric elements are present.
Note that in the case of the anisotropic plasmas, the direction of motion affects the world-sheet temperature.

To calculate the Langevin coefficients we add fluctuations in classical trailing string solutions as in \cite{Gubser:2006nz}. We choose the static gauge and consider the following form of fluctuations
\be
t=\sigma~,~~~u=\sigma~,~~~x_p=v~t+\xi(\s)+\delta x_p(\tau, \sigma)~,~~~x_{i\neq p}=\delta x_{i\neq p}(\tau, \sigma)~.
\ee
The induced metric on the world-sheet is given by $\tilde{g}_{\a\b}=g_{\a\b}+\delta g_{\a\b},$ where $\tilde{g}$ are the perturbed results.
The linear terms in fluctuations form a total derivative and we can neglect them with the particular boundary conditions. Therefore, the NG action for the fluctuations around the solution to quadratic order becomes
\bea
S_2=-\frac{1}{2\pi\alpha'}\int d\t d\s \sqrt{-g}~\frac{g^{\a\b}}{2}\,\left[N\prt{u}\rd_\a \delta x_p\,\rd_\b \delta x_p+\sum_{i\neq p}{G_{ii}}\rd_\a \delta x_i\,\rd_\b \delta x_i\right]~,
\eea
where the world-sheet determinant and the function $N\prt{u}$ are equal to
\be
g=G_{00}\,G_{uu}\,G_{pp}\,\frac{G_{00}+G_{pp}\,v^2}{G_{00}\,G_{pp}+C^2}~,\quad N(u):={\frac{G_{00}\,G_{pp}+C^2}{G_{00}+G_{pp}\,v^2}}~.
\ee
The above action can be rewritten in terms of the diagonalized metric \eq{m2} as
\bea\la{s22}
S_2&&=-\frac{1}{2\pi\alpha'}\int d\tilde{\tau} d\s \,\frac{H^{\a\b}}{2}
\left[N(u)\,\rd_\a \delta x_p\,\rd_\b \delta x_p+\sum_{i\neq p}{G_{ii}}\rd_\a \delta x_i\,\rd_\b \delta x_i\right]~,
\eea
where $H^{\a\b}=\sqrt{-h}{h}^{\a\b}$.

\subsection{Langevin Coefficients}

The quark moving with a constant velocity $v$ has similar dynamics of a Brownian motion. Its motion can be found using the generalized Langevin equations which include the components of the real-time correlation functions for the time dependent drag force. Under the assumption that for long times the time-correlation functions are proportional to $\delta$ functions, the Langevin equations become local and the diffusion coefficients are constants. The effective equation of motion takes the form
\be
\frac{dp_i}{dt}=-\eta_{D\,ij}\, p^j+\xi_i\prt{t},\label{lang1}
\ee
where $\xi_i(t)$ is the force generated by the medium, and causes the momentum broadening to the quark. In our case the background is diagonal so the friction coefficient is also a diagonal matrix. The force distribution is characterized by the two-point correlators for the longitudinal and transverse to the direction of motion $\k_a=\prt{\k_L,\k_T}$,
\be
\vev{\xi_a\prt{t}\xi_a\prt{t'}}=\k_a\d\prt{t-t'}~.
\ee
The diffusion coefficients are given by
\be
\k_{a}=\lim_{\o\rightarrow 0} G^a\prt{\omega}=-\coth\frac{\omega}{2 T_{ws}} \text{Im}G_R\prt{\o}=-2\,T_{ws}\,\lim_{\omega\rightarrow 0}\,\frac{{\rm Im}G^a_{R}(\omega)}{\omega}~.\label{kappa-a}
\ee
where $G_R$ is the anti-symmetrized retarded correlator.

A direct way to calculate the diffusion coefficients is by using the membrane paradigm \cite{Iqbal:2008by} for the world-sheet action. A fluctuation $\phi$ in the bulk of a generic theory leads to an action of the form
\be
S_2=-\frac{1}{2}\int dx du \sqrt{-g} q \prt{u}g^{ \a \b}\partial_\a \phi \partial_\b \phi~,
\ee
then the relevant transport  coefficients associated with the retarded Green function can be read from the action. It turns out that in two dimensions the metric dependence cancels out completely in the formula and the only actual dependence comes form the function $q$.

Therefore, using the effective action \eq{s22}, we obtain the transport coefficients associated to the massless fluctuations from their coupling to the effective action evaluated at the world-sheet horizon. Notice that in the case of the anisotropic plasmas, the direction of motion affects the results of the transport coefficient. So, the generic formulas for the transverse and longitudinal fluctuations and therefore the Langevin coefficients can be expressed in the background metric elements  \cite{Giataganas:2013hwa} by
\bea
\k_T=\frac{1}{\pi\alpha'}\,G_{kk}\bigg|_{u=u_0} T_{ws}~,\qquad
\k_L=\frac{1}{\pi\alpha'}\,\frac{\left(G_{00}G_{pp}\right)'} {G_{pp}\,\left(\frac{G_{00}}{G_{pp}}\right)'}\Bigg|_{u=u_0} T_{ws}~,
 \label{mpa22}
\eea
where the index $k$ denotes a particular transverse direction to that of motion $p$ and no summation is taken. The $T_{ws}$ is given in terms of metric elements by \eq{wst1}. It follows that their ratio can be written as
\be\la{ratio1}
\frac{\k_L}{\k_T}=\frac{\left(G_{00}G_{pp}\right)'} {G_{kk}G_{pp}\,\left(\frac{G_{00}}{G_{pp}}\right)'}\Bigg|_{u=u_0}~.
\ee

\subsection{Generalization of the Einstein Relation for Generic Theories}\label{subsection3}
The Einstein-like relations for motion of a quark with non-zero velocities in generic backgrounds may be also derived. The Langevin equations have the form \eq{lang1} and the linearized expressions are given by
\be
\g^3\,M_q \delta \ddot{x}_{L}=-\eta_{L}\,\delta\dot{x}_{L}+\xi_{L}~,
\quad\g \,M_q \delta \ddot{x}_{T}=-\eta_{T}\,\delta\dot{x}_{T}+\xi_{T}~,
\label{ftt1}
\ee
where the friction coefficients $\eta_{L,T}$ are defined
as
\be
\eta_a=-\lim_{\omega\rightarrow0}\frac{{\rm Im}G^a_{R}(\omega)}{\omega}~.\label{eta-a}
\ee
They are related to the coefficients $\eta_{D,a}$, by
\bea
\eta_{T}=M_q\, \gamma\,  \eta_{_{D,T}},\qquad\eta_{L}=M_q\,\gamma^3\,\left(\eta_{_{D,L}}+p\,\frac{\rd \eta_{{_D,L}}}{\rd p}\bigg|_{p=M_q v \gamma}\right)~,\label{ein2}
\eea
and therefore the broadening parameters $\k_a$  through the equation \eqref{kappa-a}, may be written as
\be
\k_{a}=2\, T_{ws}\, \eta_{a}~.\label{einstein}
\ee
We make the consistency check, and we find that for any generic background the expressions for $\k_{L}$ and $\k_{T}$ given by the equations \eq{einstein} and \eq{ein2} agree with the relevant results we have introduced in \eq{mpa22}.

In the anisotropic theories the coefficient $\eta_{{D,L}}$ given by  \eq{etaD}, is different to the  $\eta_{{D,T}}$ which can be read from \eq{einstein} and \eq{ein2}, namely
\bea
\eta_{D,T}=\frac{G_{kk}(u_0)}{2\pi \alpha' M_q \gamma}.
\eea
where $G_{kk}$ is one of the transverse metric components and might not be equal to $G_{pp}$. This is in contrast to the isotropic theories. Nevertheless, even in generic theories the diffusion and friction coefficients satisfy a version of the Einstein relations of the form
\bea
\frac{\k_T}{\eta_{D,T}}=2\, M_q\, \gamma \,T_{ws}~.\label{einstein-2}
\eea
For isotropic backgrounds this result is similar to the one obtained in \cite{HoyosBadajoz:2009pv,Gursoy:2010aa}.

\section{Anisotropic backgrounds}\label{section4}

\subsection{Space-dependent Axion Deformed Background}

The anisotropic background used here is a top-down model which is  a solution to the type IIB supergravity equations. In the dual field theory it can be though as a deformed version of the $\cN=4$ finite temperature sYM with a $\theta$ parameter term depending on the anisotropic direction  $x_3$ \cite{aniso1}.

In the gravity dual side the $\theta$ angle is related to the axion of the type IIB supergravity through the complexified coupling constant and therefore an axion with space dependence will be present in the anisotropic background. The geometry of the resulting supergravity solution has a singularity in the IR which is hidden behind the horizon and the solution can be viewed as a renormalization group flow from an isotropic UV point at the asymptotic boundary to an anisotropic IR in the near horizon limit.

In the string frame the background is given by
\bea
&&ds^2 =\frac{1}{u^2}\left( -\cF \cB\, dx_0^2+dx_1^2+dx_2^2+\cH dx_3^2 +\frac{ du^2}{\cF}\right)+ {\cal Z} \, d\Omega^2_{S^5}\,\nn \\
&&  \chi = a x_3, \qquad \phi=\phi(u) \,,
\label{metric-MT}
\eea
where $\phi$ is the dilaton, $\chi$ is the axion, and $a$ is the anisotropic parameter measured in units of inverse length. The boundary of the metric is at $u=0$ and we set the AdS radius to one. For large anisotropies the solution to the supergravity equations can be found by solving the equations of motion numerically.  An analytic form of the functions $\cF, \cB, \cH$ and $\cZ$ can be found when the anisotropy over temperature is small enough, $a/T\ll 1$.  The expansions up to second order in $a/T$ around a black D3-brane solution give
\bea\nn
&&\cF(u) = 1 - \frac{u^4}{u_h^4} + a^2 \cF_2 (u)  +\mathcal{O}(a^4)\\
&&\cB(u) = 1 + a^2 \cB_2 (u) +\mathcal{O}(a^4)~, \la{smallB}\\\nn
&&\cH(u)=e^{-\phi(u)},\quad\cZ(u)=e^{\frac{\phi(u)}{2}},\quad\mbox{where}\quad \phi(u) =  a^2 \phi_2 (u)  +\mathcal{O}(a^4)~,
\eea
with
\bea\nn
&&\cF_2(u)= \frac{1}{24 u_h^2}\left[8 u^2( u_h^2-u^2)-10 u^4\log 2 +(3 u_h^4+7u^4)\log\prt{1+\frac{u^2}{u_h^2}}\right]~,\\\la{smallF}
&&B_2(u)= -\frac{u_h^2}{24}\left[\frac{10 u^2}{u_h^2+u^2} +\log\left(1+\frac{u^2}{u_h^2}\right)\right] \,,\\\nn
&&\phi_2(u) = -\frac{u_h^2}{4}\log\prt{1+\frac{u^2}{u_h^2}}\,.
\eea
The position of the horizon $u_h$ is given in terms of temperature and the anisotropic parameter as
\be\label{uht11}
u_h=\frac{1}{\pi T}+a^2 \frac{5 \log2-2}{48 \pi^3 T^3}+\mathcal{O}(a^4)~.
\ee

The energy and the pressures of the boundary theory are defined via the expectation value of energy-momentum tensor near the boundary. The pressure along the anisotropic and the transverse space  differ, where for small anisotropies $P_{\parallel}<P_{\perp}$, while for larger ones the
inequality gets inverted.

Notice that the metric \eqref{metric-MT} is always prolate for any anisotropy, while the pressures in small anisotropies are oblate and in larger prolate.

\subsection{Bottom-Up Anisotropic Backgrounds}

In this subsection we review the anisotropic bottom-up background \cite{aniso2}. The five dimensional metric with a stationary anisotropic energy-momentum tensor satisfying $\varepsilon=2P^{\perp}+P^{\parallel}$ is given by
\be
ds^2=\frac{1}{u^2}\,\left(-a(u)\,dt^2+b(u)\, \left(dx_1^2+dx_2^2\right)+c(u)\,dx_3^2+du^2\right)~,\label{JW-metric}
\ee
where $u$ is the radial coordinate with the boundary at $u=0$. The metric functions have the form
\bea\nn
&&a(u)=\left(1+A^2\,u^4\right)^{\frac{1}{2}-\frac{1}{4} \sqrt{36-2B^2}}\,\left(1-A^2\,u^4\right)^{\frac{1}{2} +\frac{1}{4}\sqrt{36-2B^2}}\,,\\
&&b(u)=\left(1+A^2\,u^4\right)^{\frac{1}{2}+\frac{B}{6} +\frac{1}{12}\sqrt{36-2B^2}}\,\left(1-A^2\,u^4\right)^{\frac{1}{2} -\frac{B}{6}-\frac{1}{12}\sqrt{36-2B^2}}\,,\\\nn
&&c(u)=\left(1+A^2\,u^4\right)^{\frac{1}{2}-\frac{B}{3} +\frac{1}{12}\sqrt{36-2B^2}}\,\left(1-A^2\,u^4\right)^{\frac{1}{2} +\frac{B}{3}-\frac{1}{12}\sqrt{36-2B^2}}\,.
\eea
The parameters $A$ and $B$ appear to the pressures via
\bea
P^{\parallel}=\frac{A^2}{6}\,\sqrt{36-2B^2}-\frac{2}{3}\,A^2\,B~,\quad
P^{\perp}=\frac{A^2}{6}\,\sqrt{36-2B^2}+\frac{1}{3}\,A^2\,B~.
\la{JW-p}
\eea
The five dimensional static AdS black brane solution can be reached for $B=0$ or at the limit $u\rightarrow 0$.



The interesting feature of the bottom-up model is that it may have oblate and prolate geometries and pressure anisotropies, depending on the values of the parameters. Positive (negative) values for $B$ lead to oblate (prolate) geometries. In the following we will use $A=1$ and the two special values
$B=\sqrt{2}\Rightarrow P^\parallel=0$ for oblate and $B=-\sqrt{6}\Rightarrow P^\perp=0$ for prolate. Notice that this model has a mild naked singularity in the bulk, where however the definitions of infalling boundary conditions are still possible.

\section{Langevin Diffusion Coefficients in the Axion Deformed Anisotropic Theory}\label{section5}
In this section we study the Langevin diffusion coefficients in the top-down anisotropic deformed $\cN=4$ sYM. The analytical analysis is done for small anisotropies while for larger ones we use numerics.
\subsection{Small Anisotropy}
We study analytically the Langevin coefficients in the small $a/T$ limit where the metric functions  are known \eqref{smallB}. For presentation purposes we give in the Appendix \ref{app1} some of the analytical functions appearing in this section, while we note and discuss their useful properties in the main text.
Moreover, all our results in this section are up to order $\cO\left(a^4\right)$, and we mention it here to avoid carrying the symbol in all the equations.

We have two different world-sheets for a string moving along and perpendicular to the anisotropic direction and therefore two different world-sheet horizons \cite{giataganasan,Chernicoff:2012iq}, which can be found from \eqref{critical-point1}, and are of the form
\be
u_0^{\perp}= u_{0,iso}\left(1+\frac{a^2}{T^2}\tilde{u}_0^{\perp}\right)~,\qquad
u_0^{\parallel}= u_{0,iso}\prt{1+\frac{a^2}{T^2}\tilde{u}_0^{\parallel}} ~,
\ee
where $\tilde{u}$ are contributions due to anisotropy and are given analytically in Appendix \ref{app1}.
The corresponding world-sheet temperatures are obtained by using \eq{wst1}
\bea
&&
T_{ws}^{\perp}=\frac{T}{\sqrt{\gamma}}\left[1-\frac{a^2}{T^2}\frac{(1-\g) \left(4+\g-\g^2+\g^2 (\g+1) \log \left(\frac{1}{\g}+1\right)\right) }{48 \pi ^2 \gamma ^2}\right]~,\\\label{TwsMTsmall}
&&
T_{ws}^{\parallel}=\frac{T}{\sqrt{\gamma}}\left[1-\frac{a^2}{T^2}\frac{1+\g^2-2 \g^3+2 \g^2  \left(\g^2-1\right) \log \left(\frac{1}{\g}+1\right) }{48 \pi ^2 \gamma ^2}\right]~.
\eea
We observe that world-sheet temperatures can not be equal to the heat bath temperature for small anisotropic parameters $a/T$ and non-zero velocities. They become equal to each other when they become equal to the heat bath temperature and that is for zero velocity. More particularly the temperatures follow the inequality
\bea
&&T_{ws}^{\parallel}<T_{ws}^{\perp}<T~,\quad \mbox{for}\quad v>0.
\eea
The $T_{ws}$ being lower than the heat bath temperature has been observed for quark motion in the isotropic cases \cite{Gursoy:2010aa,Nakamura:2013yqa} leading to holographic "refrigerator" systems. For larger anisotropies we will notice that the inequality may be inverted.

To study the Langevin coefficients we use the notation $\k_{T,L}$ which denotes the coefficient transverse or longitudinal to the quark's motion. Moreover, we introduce the notation of the upper indices referring to the directions of the anisotropic plasma. For transverse coefficients we use the upper indices as $\k_T^{\perp,\prt{\parallel}}$ describing the effect with respect to the anisotropic direction, where the first index refers to the motion of the quark in the background, in this example in $(x_1x_2)-$plane, and the second index to the direction where the broadening happens, in this example along $x_3$. For the longitudinal components the notation is simpler and the upper index just denotes the direction of the motion of the quark to the plasma.
\newline
Using the equations \eq{mpa22}, the longitudinal broadening parameters  found to be
\bea
&&\k_L^{\perp}=\pi  \gamma ^{5/2} T^3 \sqrt{ \lambda }\left[1+\frac{a^2}{T^2}\frac{-12+\gamma  (9+2\gamma+5\gamma^2)-5 \gamma^2(1+\gamma^2) \log \left(\frac{1}{\gamma }+1\right)}{48 \pi^2 \gamma^2 }\right]~,\\
&&\k_L^{\parallel}=\pi  \gamma ^{5/2} T^3 \sqrt{ \lambda }\left[1+\frac{a^2}{T^2}\frac{-3+\g^2\prt{2 \gamma+5 +2 (-4+5\gamma^2) \log \left(\frac{1}{\gamma }+1\right)}}{48 \pi^2 \gamma^2 }\right]~.
\eea
While the transverse to the motion Langevin coefficients are
\bea
&&\k_T^{\perp,\prt{\parallel}}= \pi  \sqrt{\lambda  }\gamma^{1/2} T^3\left[1+\frac{a^2}{T^2}\frac{ -4+\gamma(1+\gamma)(3+\gamma)-\gamma^2(\gamma^2-3) \log \left(\frac{1}{\gamma }+1\right)}{48 \pi^2  \gamma ^{2}}\right]~,\\
&&\k_T^{\perp,\prt{\perp}}= \pi  \sqrt{\lambda  }\gamma^{1/2} T^3\left[1+\frac{a^2}{T^2}\frac{ -4+\gamma(1+\gamma)(3+\gamma)-\gamma^2(9+\gamma^2) \log \left(\frac{1}{\gamma }+1\right)}{48 \pi^2  \gamma ^{2}}\right]~,\\
&&\k_T^{\parallel,\prt{\perp}}=\pi  \sqrt{\lambda  }\gamma^{1/2} T^3\left[1+\,\frac{a^2}{T^2} \frac{ -1+\gamma ^2+4 \gamma ^3+2 \gamma ^2(\gamma^2-6) \log \left(\frac{1}{\gamma }+1\right)}{48 \pi^2  \gamma ^{2}}\right]~.
\eea
Notice the similarity of the expressions for the coefficients
$\k_T^{\perp,\prt{\parallel}}$ and $\k_T^{\perp,\prt{\perp}}$. Although the former is always larger than the isotropic coefficient, for large velocities they move towards the same values.

For quarks moving along the anisotropic direction the corresponding coefficient has different behavior, and is modified stronger by the anisotropy. This can be explained geometrically since the anisotropic direction of the metric is modified stronger than the transverse space and these modifications happen to carry on to the particular observable. Physically it can be interpreted that for a quark moving in an anisotropic plasma, the Langevin dynamics are strongly depending on the direction of the motion of the quark and weaker on the direction of the transverse random forces. Notice also that the coefficients $\k_T^{\perp,\prt{\perp}}$, $\k_T^{\parallel,\prt{\perp}}$ and $\k_L^{\perp}$, are lower than the isotropic result until a certain speed is reached. Similar behavior has been noticed for the drag force  for moving quarks along the transverse to anisotropic direction \cite{giataganasan,Chernicoff:2012iq}.

The longitudinal Langevin coefficients for a motion along anisotropic direction  are modified even stronger compared to the isotropic theory and it is always larger than its isotropic result. The effects on both $\k_{L}$ coefficients for motion along the anisotropic direction become larger as the velocity is increased.

\begin{figure*}[!ht]
\begin{minipage}[ht]{0.5\textwidth}
\begin{flushleft}
\centerline{\includegraphics[width=70mm]{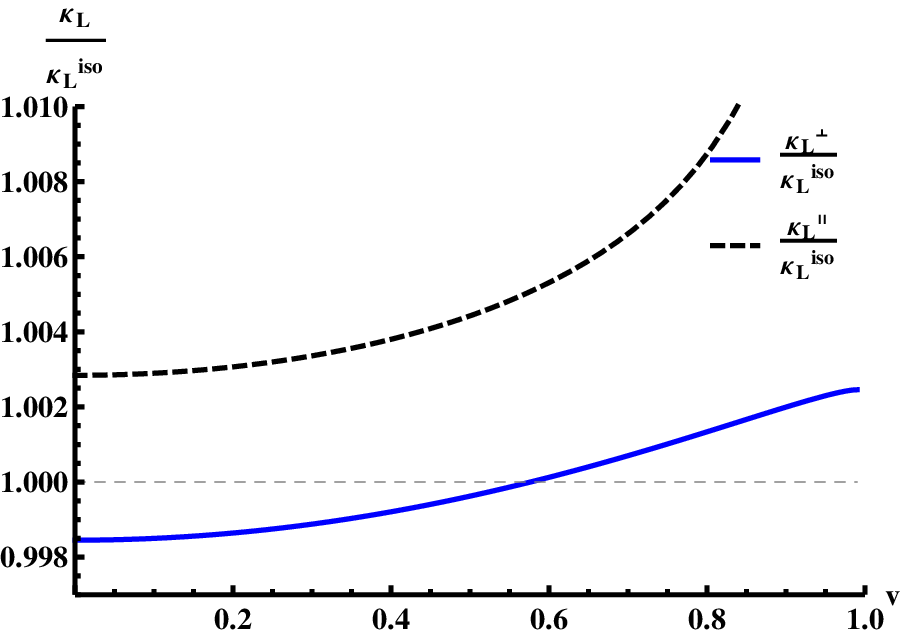}}
\caption{{\small The ratios $\k_L/\k_L^{iso}$ in terms of the velocity for different directions of motion. The plot scales has been chosen such that the crossing of a ratio to the unit is clear. Notice the strongly modified coefficient for motion along the anisotropic direction and that is always enhanced compared to the isotropic observable. Settings:  $a=0.5 T$. Smaller values of anisotropy leads to qualitatively similar results.
}}\label{fig:a1}
\end{flushleft}
\end{minipage}
\hspace{0.3cm}
\begin{minipage}[ht]{0.5\textwidth}
\begin{flushleft}
\centerline{\includegraphics[width=73mm]{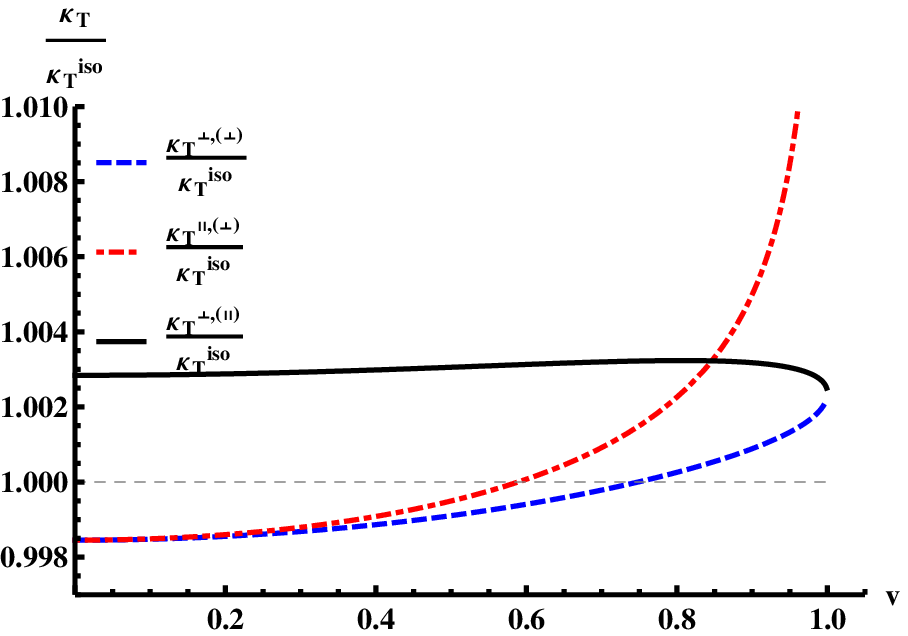}}
\caption{\small{The ratios $\k_T/\k_T^{iso}$ in terms of the velocity for different directions of motion. Notice that two of the ratios cross the unit for a particular velocity, while the other one is always larger. The strongly modified coefficient is for motion along the anisotropic direction.  Settings: as in Figure \ref{fig:a1}.}}\label{fig:a2}\vspace{1.5cm}
\end{flushleft}
\end{minipage}
\end{figure*}

These comparisons of the anisotropic Langevin coefficients to the isotropic results are depicted in Figures \ref{fig:a1} and \ref{fig:a2}.
\begin{figure*}[!ht]
\centerline{\includegraphics[width=73mm]{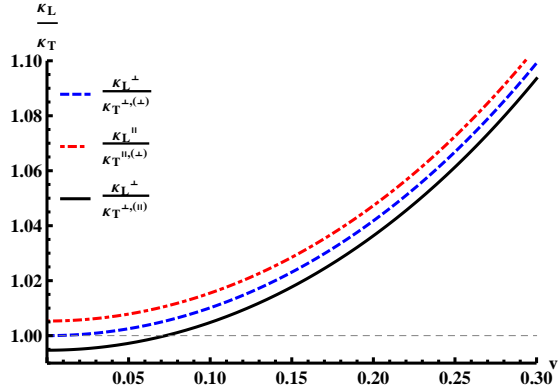}}
\caption{{ \small The rations $\k_L/\k_T$ versus the velocity. The ratio $\k_L^{\perp}/\k_T^{\perp,\prt{\parallel}}$ is the only one violating the condition $\k_L>\k_T$.
 For larger anisotropies the range of speeds that the violation happens increases. Settings:  $a=0.55 T$}}\vspace{-.5cm}
\label{fig:a3}.
\end{figure*}

An interesting remark is in order. In \cite{Giataganas:2013hwa} by finding the ratio $\k_L/\k_T$ for any theory in terms of the background metric elements, it has been noticed that the inequality $\k_L>\k_T$ holds for a large number of theories, and it has been found to get violated only in anisotropic theories. In our case this happens for a quark moving along the transverse space to anisotropy, while the transverse component of the broadening is along the anisotropic direction, Figure \ref{fig:a3}. The violation happens for small velocities, but increasing the anisotropy the range of violation may be increased to almost the whole range of the velocity.

In next section we extend our study to the large anisotropy regime of the
anisotropic axion space dependent model.

\subsection{Large Anisotropy}

The analysis of this section is done numerically, since the background in this regime is not known analytically. For large anisotropies the background metric remains prolate while the pressure anisotropies may be prolate or oblate. Moreover, in the numerical analysis we have a larger range to vary our parameters and we observe new behaviors in our observables.

\begin{figure*}[!ht]
\begin{minipage}[ht]{0.5\textwidth}
\begin{flushleft}
\centerline{\includegraphics[width=70mm]{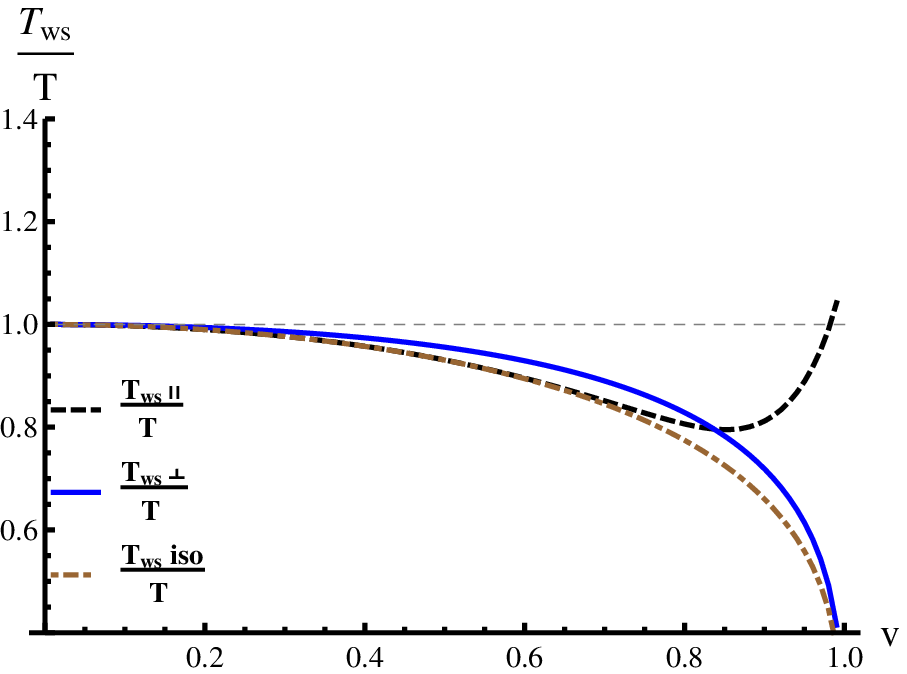}}
\caption{\small{The ratios $T_{ws}/T$ as functions of velocity for moving quarks along different directions. Notice that for large velocities the $T_{ws}^\parallel$ becomes larger than the heat bath temperature. Settings: $a/T=10$.}}\label{fig:b05}
\end{flushleft}
\end{minipage}
\hspace{0.3cm}
\begin{minipage}[ht]{0.5\textwidth}
\begin{flushleft}
\centerline{\includegraphics[width=70mm]{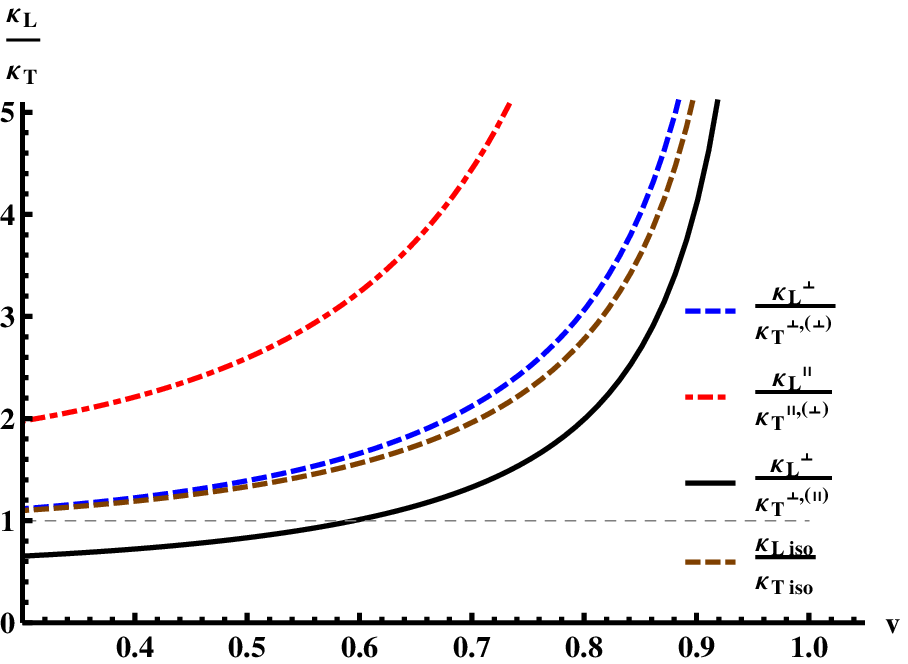}}
\caption{{\small The ratios $\k_L/\k_{T}$ as functions of velocity for different directions of motion.  Notice the larger range of velocities compared to the small anisotropy case for which $\k_L^{\perp}<\k_{T}^{\perp,(\parallel)}$. Settings: As in Figure \ref{fig:b05}.
}}\label{fig:b03}
\end{flushleft}
\end{minipage}
\end{figure*}

\begin{figure}
\begin{minipage}[ht]{0.5\textwidth}
\begin{flushleft}
\centerline{\includegraphics[width=70mm]{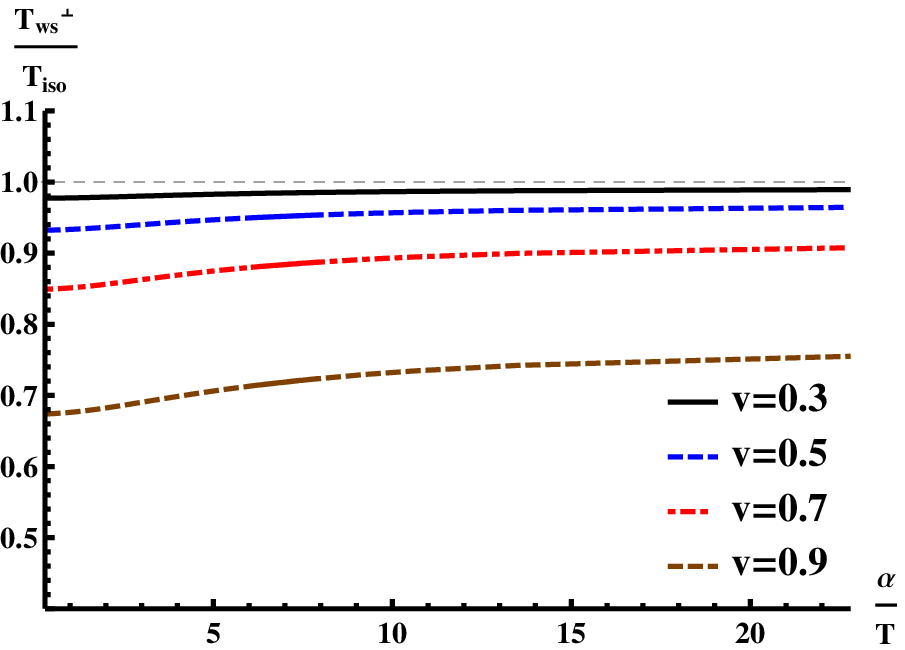}}
\caption{\small{The ratios $T_{ws}/T$ as functions of anisotropy for moving quarks along the transverse plane. Even for large velocities and anisotropies the $T_{ws}^\perp$ is lower than the heat bath temperature, as in the isotropic theories.}}\label{fig:b08}\vspace{.5cm}
\end{flushleft}
\end{minipage}
\hspace{0.3cm}
\begin{minipage}[ht]{0.5\textwidth}
\begin{flushleft}
\centerline{\includegraphics[width=70mm]{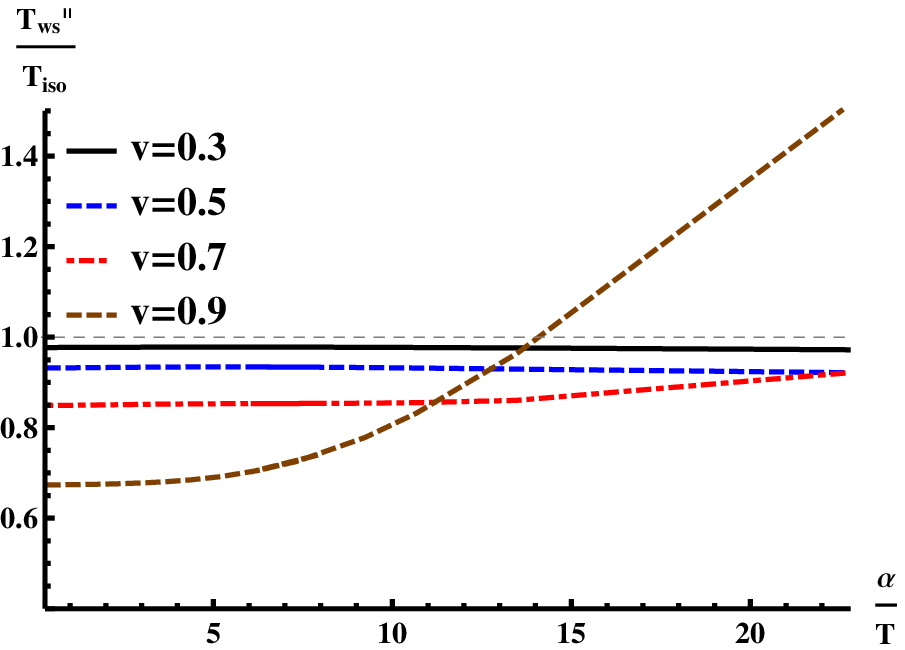}}
\caption{{\small The ratios $T_{ws}/T$ as functions of anisotropy for motion along the anisotropic direction.  Notice that for large anisotropies the $T_{ws}^\parallel$ becomes bigger than the heat bath temperature, and as velocity increases the anisotropy for this to happen gets lower.
}}\label{fig:b09}
\end{flushleft}
\end{minipage}
\end{figure}

A new observation in large anisotropies is that the effective temperature may be larger than the heat bath temperature, in contrast to the isotropic theories. For anisotropic parameters of the order $a/T= 10$ this is barely happens for very large velocities (Figure \ref{fig:b05}). As we increase the anisotropies however, the inequality $T\leq T_{ws}^\parallel$ is satisfied for larger range of speeds  as can be seen in Figure \ref{fig:b09}. This is a unique result for quarks moving in anisotropic theories. Therefore, a quark can be moving in the anisotropic theory with such a speed that the effective temperature the quark measures, is equal to the heat bath temperature. On the other hand the effective temperature for quark motion in the transverse plane is always lower than the heat bath temperature (Figure \ref{fig:b08}).

For larger velocities the world-sheet temperature $T_{ws}^\perp$ is decreasing, while the  $T_{ws}^\parallel$ is also
decreased until some specific value of anisotropies after which the effective temperature is increasing, as can be seen in  Figure \ref{fig:b09}. This is also a unique phenomenon of the anisotropic theories.
In summary we see that while the world-sheet temperature  $T_{ws}^\perp$ for quark's motion along the transverse to anisotropy direction has many common characteristics to the isotropic theories, the temperature $T_{ws}^\parallel$ for a quark motion along the anisotropic direction has very different properties.
\begin{figure*}[!ht]
\begin{minipage}[ht]{0.5\textwidth}
\begin{flushleft}
\centerline{\includegraphics[width=70mm]{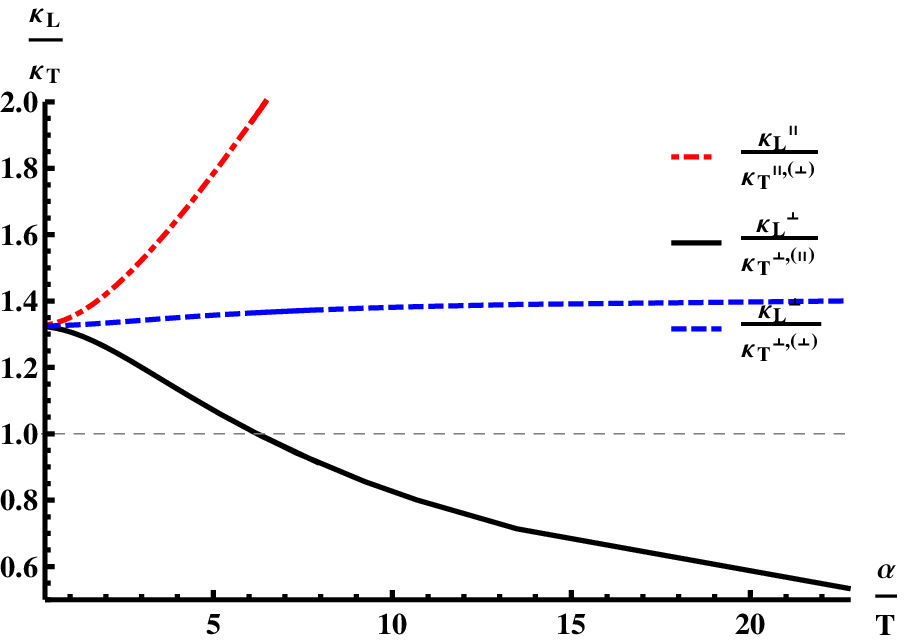}}
\caption{{\small The ratios $\k_L/\k_{T}$ depending on the anisotropy for $v=0.5$.  Notice that for anisotropic parameters $a/T\gtrsim 6$ the   inequality $\k_L^{\perp}>\k_{T}^{\perp,(\parallel)}$ does not hold.
}}\label{fig:b077}
\end{flushleft}
\end{minipage}
\hspace{0.3cm}
\begin{minipage}[ht]{0.5\textwidth}
\begin{flushleft}
\centerline{\includegraphics[width=70mm]{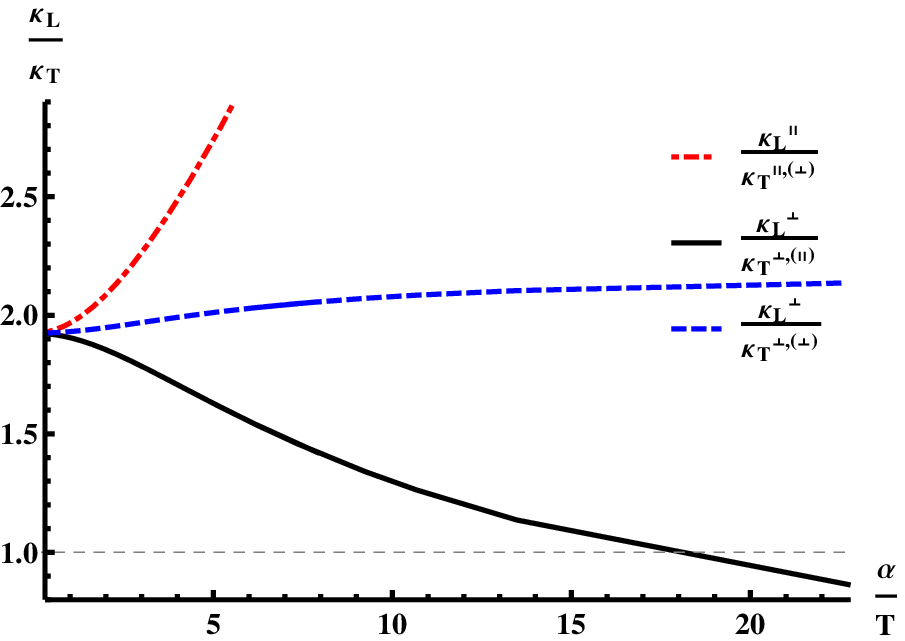}}
\caption{{\small The ratios $\k_L/\k_{T}$ depending on the anisotropy for $v=0.7$.  The increase of the velocity requires larger anisotropies where the inequality $\k_L^{\perp}>\k_{T}^{\perp,(\parallel)}$ is not satisfied.
}}\label{fig:b07}
\end{flushleft}
\end{minipage}
\end{figure*}

The behavior of the diffusion coefficients in large anisotropies is similar to that of the small anisotropies as depicted in Figures \ref{fig:a1}, \ref{fig:a2} and \ref{fig:a3}. Therefore, the qualitative explanations and the interpretations are similar to the ones given for motion of the quark in small anisotropies in the previous section. The only quantitative difference is that the effects of anisotropy in the plotted quantities become larger. Nevertheless, it is interesting to note that the ratio $\k_L^\perp/\k_T^{\perp,\prt{\parallel}}$ is lower than the unit for large range of velocities for increased anisotropies. In Figure \ref{fig:b03} we find that the ratio is lower than the unit for velocities $v\lesssim 0.6$, for an anisotropy $a/T=10$. The range of velocities that the ratio remains lower than the unit increases as the anisotropy is increased. This can be seen in Figures \ref{fig:b077} and \ref{fig:b07}, where large velocities eventually violate the inequality $\k_L^\perp>\k_T^{\perp,\prt{\parallel}}$, only requiring to have stronger anisotropic backgrounds.

\section{Langevin Diffusion Coefficients in Bottom-up Model}\label{section6}

In this section we examine the bottom-up model \eqref{JW-metric} for the oblate and prolate geometries and investigate the dependence of the coefficients on the type of the geometry.  For prolate geometry we choose the value $B=-\sqrt{6}$, for the oblate the value $B=\sqrt{2}$, while the isotropic case is for $B=0$. The parameter $A$ is chosen to be equal to the unit. Notice that the background singularity, although it is mild, affects our results especially at low velocities. Therefore we avoid to examine in detail the observables at very low speeds and we focus mostly on higher speeds.

\subsection{Oblate and Prolate Geometries}

Oblate pressure anisotropies are expected in the observed QGP, and we choose for our background the one that makes the anisotropies maximal, $B=\sqrt{2}$. The analysis can be done by applying our formulas, where solving the \eq{critical-point1} we find the world-sheet horizons $u_{0,obl}^{\perp}$, $u_{0,obl}^{\parallel}$ given  explicitly in Appendix \ref{app:j} by \eq{ob1}. The corresponding temperatures $T_{ws,obl}^{\perp}$, $T_{ws,obl}^{\parallel}$ can be found by using the equation \eq{wst1}, and given in the \eq{ob2}.

For the prolate geometry with $B=-\sqrt{6}$, analytical results can also be obtained where the world-sheet horizons $u_{0,pro}^{\perp}$, $u_{0,pro}^{\parallel}$  and the corresponding temperatures $T_{ws,pro}^{\perp}$, $T_{ws,pro}^{\parallel}$ are  given by the equations \eq{pro1} and \eq{pro2}.
Since we have chosen $A=1$ for the anisotropic case, we use it in the isotropic background and give the effective temperature and corresponding Langevin coefficients in \eqref{iso-jw}.

Our main aim using the bottom-up model is to study the dependence of the Langevin coefficients on the type of the geometry of the background and we focus on larger velocities. We look at large velocities since at low velocities the worldsheet horizon moves close to the singularity of the space, where our results are affected strongly by the singularity and can not be trusted.

\begin{figure*}[!ht]
\begin{minipage}[ht]{0.5\textwidth}
\begin{flushleft}
\centerline{\includegraphics[width=70mm]{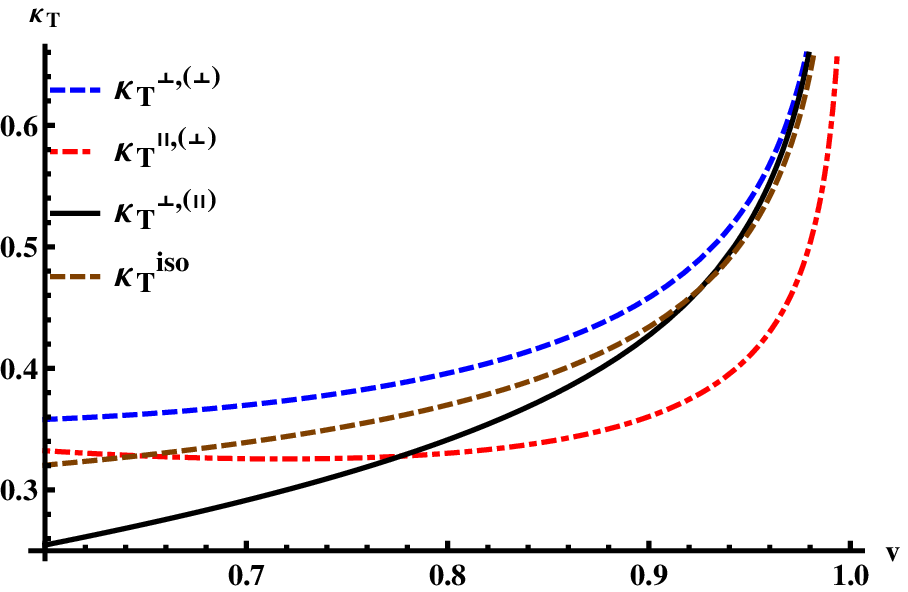}}
\caption{{\small The coefficients $\k_T$ for the oblate geometry in terms of the velocity for different directions of motion.
}}\label{fig:a11}\vspace{.5cm}
\end{flushleft}
\end{minipage}
\hspace{0.3cm}
\begin{minipage}[ht]{0.5\textwidth}
\begin{flushleft}
\centerline{\includegraphics[width=70mm]{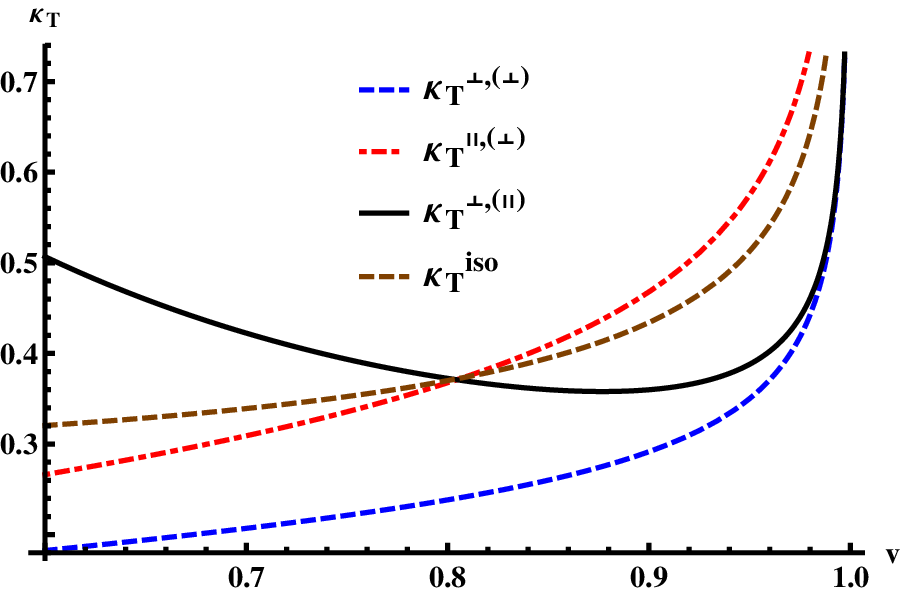}}
\caption{\small{ The coefficients $\k_T$ for the prolate geometries in terms of the velocity for different directions of motion. Notice the qualitative interchange of behaviors compared to the oblate geometries.}}\label{fig:a12}
\end{flushleft}
\end{minipage}
\end{figure*}

\begin{figure*}[!ht]
\begin{minipage}[ht]{0.5\textwidth}
\begin{flushleft}
\centerline{\includegraphics[width=70mm]{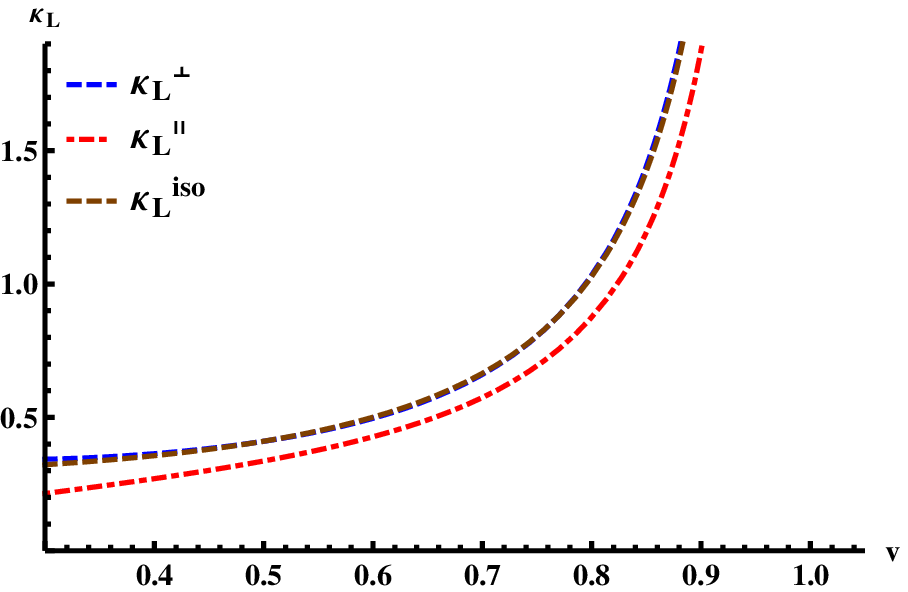}}
\caption{{\small The longitudinal Langevin coefficients  $\k_L$ in terms of the velocity for different directions of motion for oblate geometries. Notice the decreased quantity for motion along the anisotropic direction, while the transverse component almost coincides compared to the isotropic results.
}}\label{fig:a14}
\end{flushleft}
\end{minipage}
\hspace{0.3cm}
\begin{minipage}[ht]{0.5\textwidth}
\begin{flushleft}
\centerline{\includegraphics[width=70mm]{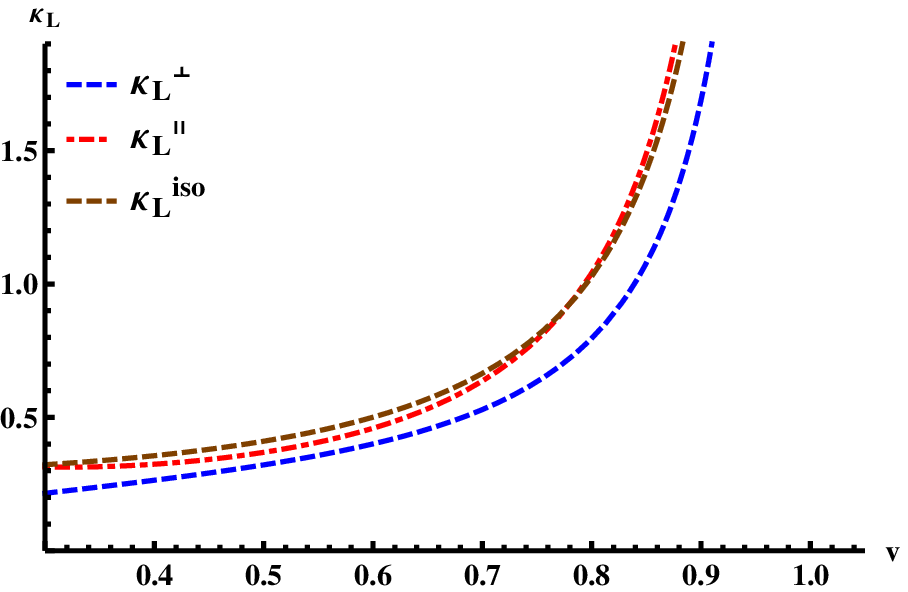}}
\caption{\small{The longitudinal Langevin coefficients  $\k_L$ in terms of the velocity for different directions of motion for prolate geometries. From oblate to prolate geometries the $\k_L$ coefficients along anisotropic and transverse to the anisotropic directions interchange qualitative behaviors. }}\label{fig:a15}
\end{flushleft}
\end{minipage}
\end{figure*}

The Langevin coefficients can be calculated analytically for both type of geometries by applying our formulas \eq{mpa22}. We give the results in  Appendix \ref{app:j} and we plot the individual coefficients below. We find that when the geometry changes from oblate to prolate, the behavior of the Langevin coefficients at large velocities it is almost inverted compared to the isotropic coefficients. Moreover, the behavior of $\k_{T}^{\parallel,\prt{\perp}}$ is qualitatively interchanged with that of $\k_{T}^{\perp,\prt{\parallel}}$ when going from prolate to oblate geometries (Figures \eqref{fig:a11} and \eqref{fig:a12}). This is even more obvious in longitudinal components of the Langevin coefficients where $\k_{L}^{\parallel}$  and $\k_{L}^{\perp}$ interchange qualitative behaviors (Figures \eqref{fig:a14} and \eqref{fig:a15}).
This can be understood geometrically since the long axis of the ellipsoid changes direction when going from oblate to prolate geometries.
The effect of the singularity for this qualitative picture seems to be not significant and therefore we conclude that the noise factors are crucially affected by the type of the geometry. In fact some of them interchange qualitative behavior when going from one to the other type of the geometry. There is also a hint from the metric of the space for this behavior, since the transverse and the longitudinal metric elements almost interchange their form as the type of geometry changes.

From a geometrical point of view we may also explain the rest of the behavior of the coefficients. The larger the speed, the closer world-sheet horizon to the boundary, where the evaluations of the quantities is done, and the anisotropic effects on the metric may be weaker.  For speeds $v\rightarrow 1$, the world-sheet horizon is very close to the background boundary and the anisotropic effects in the geometry are minimal.
Therefore,  a moving quark in the dual anisotropic plasma moving with extremely large velocities, has minor contributions of the anisotropy to its noise coefficients. However for relatively large speeds but away of the speed of light, although the world-sheet horizon moves towards the boundary, the Langevin coefficients still capture the anisotropic contributions.

\begin{figure*}[!ht]
\begin{minipage}[ht]{0.5\textwidth}
\begin{flushleft}
\centerline{\includegraphics[width=70mm]{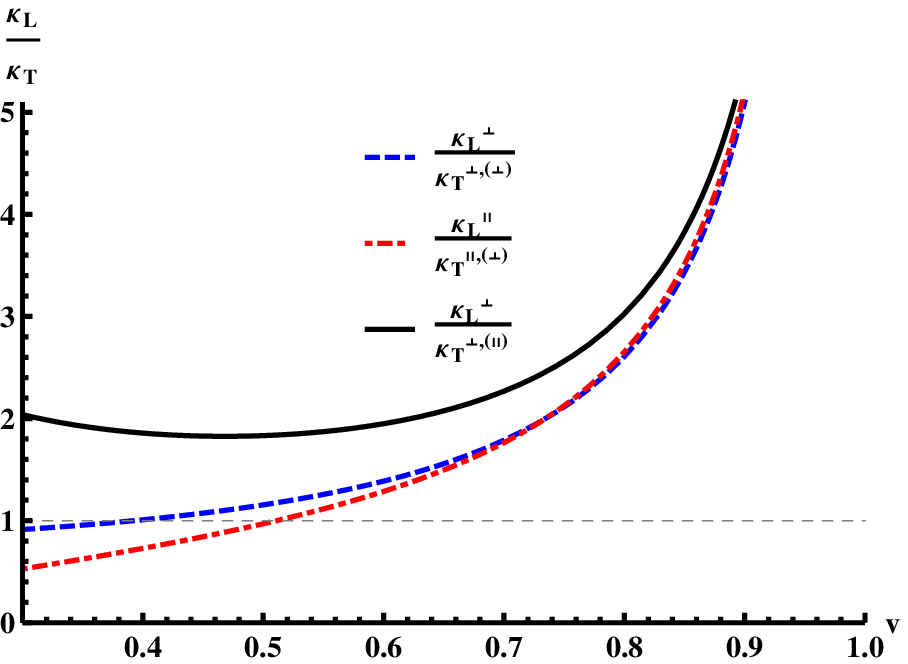}}
\caption{{\small The ratios $\k_L/\k_{T}$ depending on the velocity for different directions of motion for oblate geometries. Two of the ratios are lower than the unit for intermediate velocities
}}\label{fig:a4}
\end{flushleft}
\end{minipage}
\hspace{0.3cm}
\begin{minipage}[ht]{0.5\textwidth}
\begin{flushleft}
\centerline{\includegraphics[width=70mm]{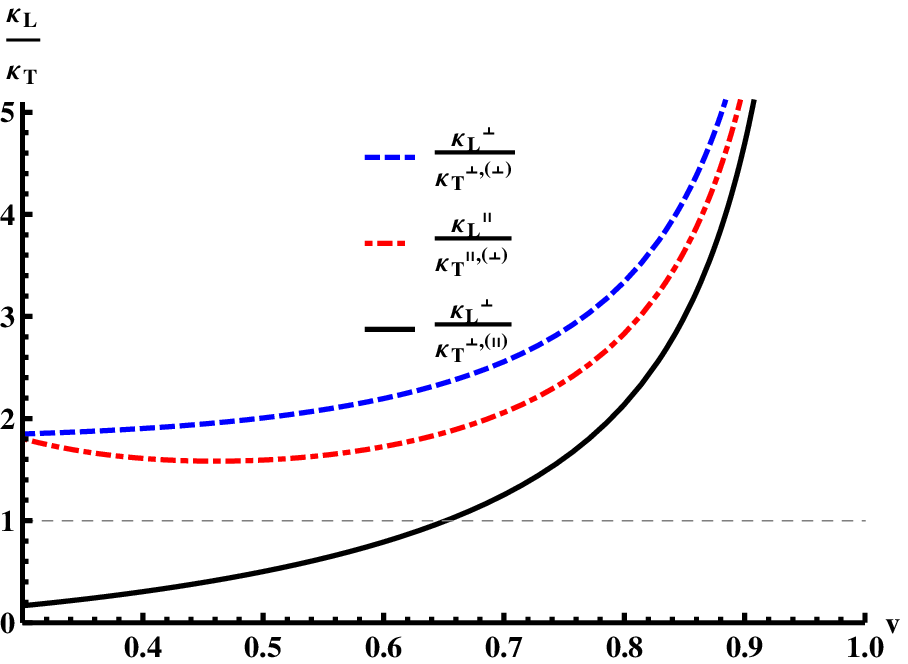}}
\caption{\small{The ratios $\k_L/\k_{T}$ depending on the velocity for different directions of motion for the prolate geometries. One ratio is smaller than the unit, and is the one that is always larger for the prolate geometries. }}\label{fig:a5}
\end{flushleft}
\end{minipage}
\end{figure*}

Notice that for prolate geometries we find in general, large deviations of the coefficients for motion along the anisotropic direction. This has been also observed in the axion space-dependent anisotropic model. Therefore quarks that are moving along the beam direction would feel the anisotropic effects in their noise factors much stronger than moving in the transverse space. This has been also observed in several other observables \cite{giataganasan}.

The expressions of the ratios of the longitudinal and transverse diffusion coefficients are simplified. For the oblate geometries read
\bea
&&\frac{\k_{L,obl}^{\perp}}{\k_{T,obl}^{\perp,\prt{\perp}}}= \frac{1+\sqrt{2}+\left(\sqrt{2}-1\right) v^{\frac{4 \sqrt{2}}{3}}}{3 \left(1-v^{\frac{4 \sqrt{2}}{3}}\right)}    ~,\qquad
\frac{\k_{L,obl}^{\perp}}{\k_{T,obl}^{\perp,\prt{\parallel}}}= \frac{\sqrt{2}+1+\left(\sqrt{2}-1\right) v^{\frac{4 \sqrt{2}}{3}}}{3 v^{2/3} \left(1-v^{\frac{4 \sqrt{2}}{3}}\right)}~,\nn\\
&&\frac{\k_{L,obl}^{\parallel}}{\k_{T,obl}^{\parallel,\prt{\perp}}}= \frac{v \left(1+\sqrt{2}-\left(\sqrt{2}-1\right) v^{2 \sqrt{2}}\right)}{\sqrt{2} \left(1-v^{2 \sqrt{2}}\right)}  ~,
\eea
while for the prolate geometries the ratios of the coefficients are
\bea
&&\frac{\k_{L,pro}^{\perp}}{\k_{T,pro}^{\perp,\prt{\perp}}}=  \frac{2+\sqrt{6}+\left(2-\sqrt{6}\right) v^{4 \sqrt{\frac{2}{3}}}}{\sqrt{6} \left(1-v^{4 \sqrt{\frac{2}{3}}}\right)}   ~,\qquad
\frac{\k_{L,pro}^{\perp}}{\k_{T,pro}^{\perp,\prt{\parallel}}}=\frac{v^2 \left(2+\sqrt{6}+\left(2-\sqrt{6}\right) v^{4 \sqrt{\frac{2}{3}}}\right)}{\sqrt{6} \left(1-v^{4 \sqrt{\frac{2}{3}}}\right)}~,\nn\\
&&\frac{\k_{L,pro}^{\parallel}}{\k_{T,pro}^{\parallel,\prt{\perp}}}= \frac{1+v^{2\sqrt{\frac23}}}{\sqrt{6}\,v\,\left(1-v^{2\sqrt{\frac23}}\right)}  ~.
\eea
The inequality $\k_L>\k_T$ is violated in the oblate geometries for the fractions $\k_{L,obl}^{\perp}/\k_{T,obl}^{\perp,\prt{\perp}}$ and
$\k_{L,obl}^{\parallel}/\k_{T,obl}^{\parallel,\prt{\perp}}$, while
for the prolate geometries for the  ratio $\k_{L,pro}^{\perp}/\k_{T,pro}^{\perp,\prt{\parallel}}$, i.e. Figures \eqref{fig:a4} and \eqref{fig:a5}. This is one more particular example where the violation of the universal relation happens for the anisotropic backgrounds  \cite{Giataganas:2013hwa}.

Notice that we have not seen negative excess noise in this anisotropic background using the conditions derived in \cite{Giataganas:2013hwa}. We find positive excess noise for the quark's motion with any velocity and along any direction, in the two extreme oblate and prolate backgrounds.

\section{Common Results Between the two Models}\label{section7}

In this section we report some further common results  between the bottom-up and the top-down anisotropic models studied in the paper. We study the dependence of the Langevin coefficients on the type of geometry by fixing the ratio of pressures, and we notice that the only clear qualitative similarity between the two models, is in terms of the ratios $\k_L/\k_T$. We focus only in the prolate regime since the axion deformed model has only this type of the geometry.

A way to relate the two models is to use the coefficient $\Delta$
\be
\Delta:=\frac{P^{\perp}}{P^{\parallel}}-1~,
\ee
which measures the degree of pressure anisotropy and depends on the anisotropic parameters. For low values of $\Delta$ an analytic relation may be found with the parameter $a$ of the axion deformed anisotropic theory \cite{giataganasan}
\be
\D\simeq \frac{a^2}{2\pi^2 T^2}
\ee
and with the parameter $B$ in the bottom-up model \cite{Rebhan:2011ke, Giataganas:2013lga}
\be
\D\simeq B~.
\ee
for low anisotropies. For larger anisotropies using the same techniques, the computation may be done numerically.

In order to understand the comparison we need to investigate the type of geometries associated to each value of $\D$. In the axion deformed anisotropic model the geometry is always prolate, while for low values of $a/T$ the pressure anisotropy is oblate, and for larger it becomes and remains prolate. In the bottom-up model, when the pressure anisotropy becomes prolate(oblate) the geometry is also prolate(oblate).

It would be natural to expect that if there is any correlation of the observables between these two models, it is more likely to be between the same type of geometries. By fixing the pressure anisotropies as in table \ref{table-1}\footnote{ Where  $\tilde{\phi}_h=\phi(u_h)+\frac{4}{7} \log a$, are the parameters for the solutions of the axion deformed model and $B$ is the one of the bottom up.} we observe a similarity between the behavior of the ratios $\k_L/\k_T$ of these two models while the individual behavior of the noise coefficients does not show any other clear similarity.
\begin{table}[ht]
\caption{Background Parameters }
\centering
\begin{tabular}{c c c c c c}
\hline \hline
$\tilde{\phi}_h$ & $u_h$ & $a/T$ & $T$ & $\Delta$& $B$  \\
\hline
 0.06 & 1.07 & 6.43  & 0.318 & -1.00  & -2.45  \\
\hline
\end{tabular}\label{table-1}
\end{table}

\begin{figure*}[!ht]
\begin{minipage}[ht]{0.5\textwidth}
\begin{flushleft}
\centerline{\includegraphics[width=70mm]{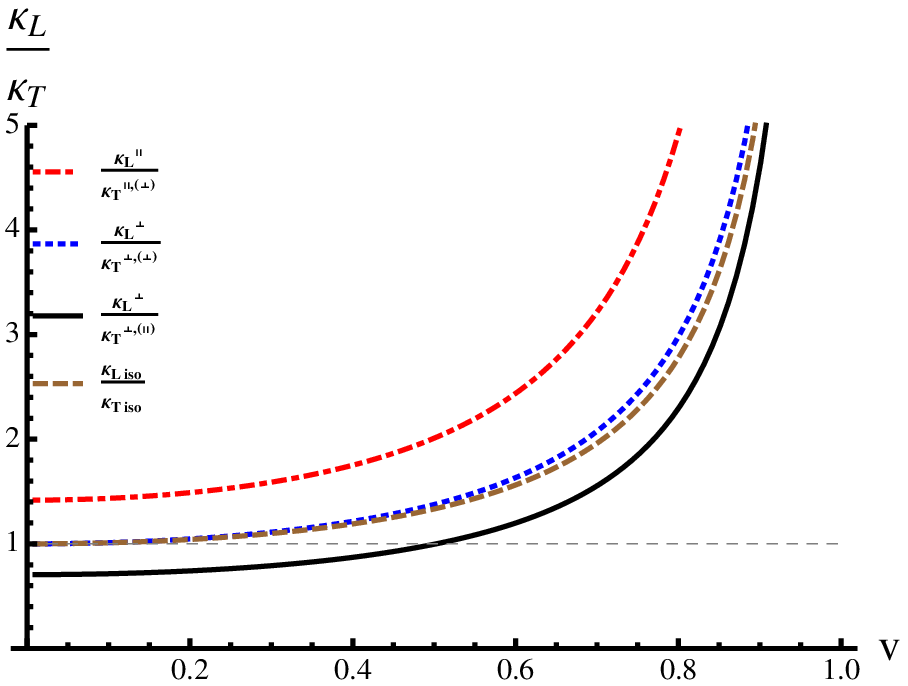}}
\caption{\small{The ratios $\k_{_L}/\k_{_T}$   for $\Delta=-1.00$ axion deformed anisotropic model. The corresponding anisotropic parameter is $a/T\sim 6.43$, where geometry and pressure anisotropies are oblate.}}\label{fig:b0m}
\end{flushleft}
\end{minipage}
\hspace{0.3cm}
\begin{minipage}[ht]{0.5\textwidth}
\begin{flushleft}
\centerline{\includegraphics[width=70mm]{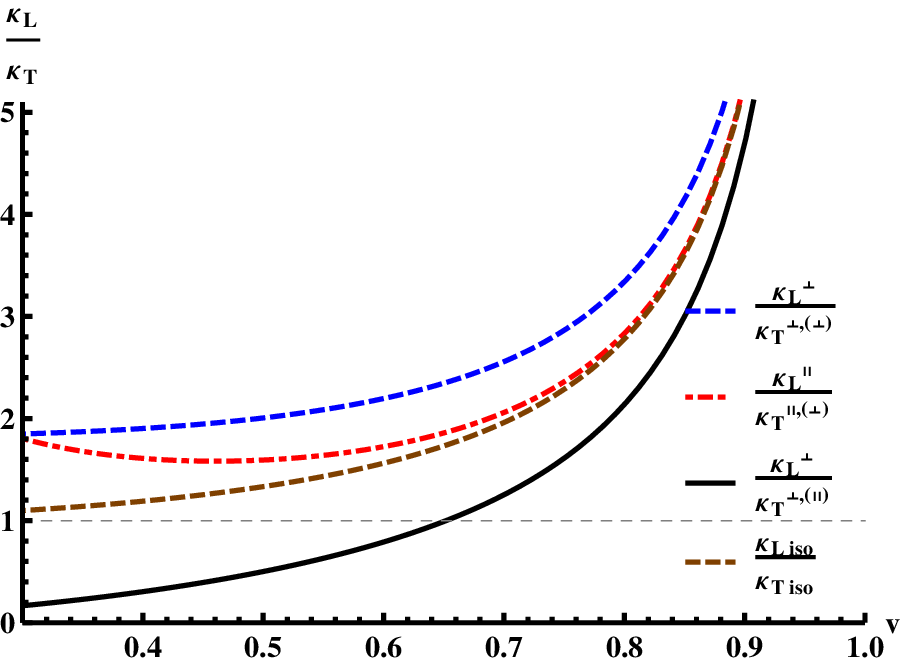}}
\caption{{\small The ratios $\k_{_L}/\k_{_T}$  for $\Delta=-1.00$ corresponding to anisotropic parameter $B\sim -2.45$ in the bottom up model.
}}\label{fig:b01}
\end{flushleft}
\end{minipage}
\end{figure*}

In Figures \ref{fig:b0m} and \ref{fig:b01} we plot the ratios $\k_L/\k_T$ for $\D=-1$ corresponding to prolate geometries and pressure anisotropies. The only ratio that is smaller than the unit in both geometries is the $\k_L^\perp/\k_T^{\perp,\prt{\parallel}}$, and has a crossing for quark speed around $v\simeq 0.5$ and $v\simeq 0.6$ in the two models. The isotropic ratio is the next larger one. But the ratios $\k_L^\perp/\k_T^{\perp,(\perp)}$ and $\k_L^\parallel/\k_T^{\parallel,(\perp)}$ come with different ordering in the two plots.

Therefore, when we compare prolate geometries  between the two anisotropic models, we see clear qualitative similarities only for some of the ratios $\k_L/\k_T$ and not the individual coefficients.

\section{Summary and discussion}\label{section8}

In this paper we have studied the Langevin diffusion coefficients in strongly coupled anisotropic plasmas. We have studied the coefficients in the top-down model of the space dependent axion deformed theory which has prolate geometry and oblate or prolate pressure anisotropy. We have found the dependence of the Langevin coefficients to the anisotropy and have compared them to each other and to the isotropic results studying extensively the effect of the anisotropies to the coefficients. Moreover, several new interesting features appear in the anisotropic theories. We show that for large anisotropies the world-sheet temperature $T_{ws}^\parallel$ for motion along the anisotropic direction may be larger or lower than the heat bath temperature. This is in contrast to the isotropic theories, where the effective temperature of the quark is always lower than the heat bath temperature. In the anisotropic theories, the effective temperature depends strongly on the velocity and anisotropy. For larger anisotropies the world-sheet temperature becomes larger than the heat bath temperature for lower velocities. We have also examined the inequality $\k_L>\k_T$, known to be true for large number of isotropic theories and to be violated in anisotropic theories according to \cite{Giataganas:2013hwa}. We show that increase of the anisotropy leads to increased range of speeds where the inequality is being violated.

Moreover we observe that motion along the anisotropic direction affects
the Langevin coefficients stronger. This is not unexpected, and reflects to an extend the degree of modification of the anisotropy to the  metric elements, which is stronger along the anisotropic direction. This has been also noticed for several other observables \cite{giataganasan}.

In order to study the qualitative dependence of the Langevin coefficients to the type of the geometry we use a bottom-up anisotropic model, which allows both prolate and oblate geometries. Focusing on larger velocities, we find that indeed the Langevin coefficients are affected by the type of the geometry and certain coefficients interchange each other qualitative behaviors. This interchange is very clear for $\k_L^\parallel$ and  $\k_L^\perp$, when going from oblate to prolate geometries. This is not unexpected since the large axis of the ellipsoid is rotated with the change of the geometry and the relevant metrics elements also interchange qualitative behaviors. In this model we have also found that the inequality $\k_L>\k_T$ is modified for any type of geometry. By fixing the anisotropy between the two top-down and bottom-up models we observe that a clear similarity between them is when comparing for the prolate geometries the ratio $\k_L^\perp/\k_T^{\perp,\parallel}$, which is lower than the unit until a certain speed is reached.

It is particularly interesting that we have not found negative excess noise in any of these two anisotropic models by examining the conditions of \cite{Giataganas:2013hwa}. So far there is no known anisotropic dual plasma that a quark's  motion  leads to negative excess noise and it would be very interesting to find a particular consistent anisotropic theory that this is allowed.

\medskip

\noindent{\bf Acknowledgments.} We would like to thank G. Georgiou, U. Gursoy, N. Irges and R. Janik for useful discussions. D.G. would like to thank the NTUA for hospitality. H. S.  would like to thank ICTP and Simons Summer Workshop 2013 in Mathematics and Physics, where this work was partly done for hospitality. The research of D.G is implemented under the "ARISTEIA" action of the "operational programme education and lifelong learning" and is co-funded by the European Social Fund (ESF) and National Resources. H.S. is supported by the South African National Research Foundation (NRF) and Foundation for Polish Science MPD Programme co-financed by the European Regional Development Fund, agreement NO. MPD/2009/6.


\appendix
\section{Analytic Results for Small Anisotropy}\label{app1}
The world-sheet horizons for a moving string along and perpendicular to the anisotropy in small anisotropy limit of axion deformed theory are given by
\be
u_0^{\parallel}= \frac{1}{\pi\,T\,\sqrt{\gamma}}\prt{1+\frac{a^2}{T^2}\tilde{u}_0^{\parallel}} ~,\quad
u_0^{\perp}= \frac{1}{\pi\,T\,\sqrt{\gamma}} \left(1+\frac{a^2}{T^2}\tilde{u}_0^{\perp}\right)~,
\ee
where
\bea
\tilde{u_0}^{\perp}=\frac{\left(4+\gamma ^2\right) \log \left(\frac{1}{\gamma}+1\right)-\gamma -1}{48 \pi ^2}\,,\quad
\tilde{u_0}^{\parallel} =\frac{\left(7-2\gamma ^2\right) \log \left(\frac{1}{\gamma}+1\right)-\gamma -1}{48 \pi ^2}~.
\eea
The ratios for $\k_L/\k_T$ are
\bea
&&
\frac{\k_L^{\parallel}}{\k_T^{\parallel,\prt{\perp}}}=\gamma^2+ \frac{a^2}{T^2}\frac{-1-\gamma^2(-2+\gamma) +2 \gamma ^2 \left(1+2 \gamma ^2\right)  \log \left(\frac{1}{\gamma }+1\right)}{24 \pi ^2}~,\\
&&\frac{\k_L^{\perp}}{\k_T^{\perp,\prt{\perp}}}=\gamma ^2+\frac{a^2}{T^2}\frac{(\gamma-1)(4+\gamma+2\gamma^2) +2 \gamma ^2 \left(1-\gamma ^2\right)  \log \left(\frac{1}{\gamma }+1\right)}{24 \pi ^2}~,\\
&&\frac{\k_L^{\perp}}{\k_T^{\perp,\prt{\parallel}}}=\gamma ^2+\frac{a^2}{T^2}\frac{(\gamma-1)(4+\gamma+2\gamma^2) -2 \gamma ^2 \left(\gamma ^2+1\right)  \log \left(\frac{1}{\gamma }+1\right)}{24 \pi ^2}~.
\eea
\section{Oblate, Prolate and Isotropic Geometries Analytic Expressions}\la{app:j}
The world-sheet horizon for quarks moving along and transverse to anisotropy for oblate geometries are given respectively by
\be\la{ob1}
u_{0,obl}^{\parallel}=\frac{\left(1-v^{\sqrt{2}}\right)^{1/4}}{\left(1+v^{\sqrt{2}}\right)^{1/4}}~,\qquad u_{0,obl}^{\perp}=\frac{\left(1-v^{\frac{2\sqrt{2}}{3}}\right)^{1/4}}{\left(1+v^{\frac{2\sqrt{2}}{3}}\right)^{1/4}}~,
\ee
while the corresponding world-sheet temperatures are
\bea\la{ob2}
&&T_{ws,obl}^{\parallel}=\frac{v^{1-\frac{3}{2 \sqrt{2}}} \left(1-v^{2 \sqrt{2}}\right)^{1/4} \left(1+\sqrt{2}-\left(\sqrt{2}-1\right) v^{2 \sqrt{2}}\right)^{1/2}}{{2^{1/4}} \pi }~,\\ &&T_{ws,obl}^{\perp}=\frac{\sqrt{3}\, v^{\frac{2}{3}-\frac{\sqrt{2}}{2}}\prt{1-v^{\frac{4 \sqrt{2}}{3}}}^{1/4} \left(\sqrt{2}+2+\left(2-\sqrt{2}\right) v^{\frac{4 \sqrt{2}}{3}}\right)^{1/2}}{2^{5/4} \pi} ~.
\eea
For a quark moving in prolate background we find the following world-sheet horizons
\be\la{pro1}
u_{0,pro}^{\parallel}=\frac{\left(1-v^{\sqrt{\frac{2}{3}}}\right)^{1/4}}{\left(1+v^{\sqrt{\frac{2}{3}}}\right)^{1/4}}~,\qquad
u_{0,pro}^{\perp}=\frac{\left(1-v^{2\sqrt{\frac{2}{3}}}\right)^{1/4}} {\left(1+v^{2\sqrt{\frac{2}{3}}}\right)^{1/4}}~,
\ee
and the world-sheet temperatures
\bea\la{pro2}
\begin{array}{l}
T_{ws,pro}^{\perp} =\frac{ v^{1-\sqrt{\frac{3}{2}}} {\left(1-v^{4 \sqrt{\frac{2}{3}}}\right)^{1/4}} {\left(3\sqrt{2}+2\sqrt{3}+\left(2-\sqrt{3}-3\sqrt{2}\right) v^{4 \sqrt{\frac{2}{3}}}\right)^{1/2}}}{2^{5/4} \pi }\,,\\\\
T_{ws,pro}^{\parallel}=\frac{\left(\frac{3}{2}\right)^{1/4} v^{\frac{1}{4} \left(2-\sqrt{6}\right)}\, {\left(1-v^{4 \sqrt{\frac{2}{3}}}\right)^{1/2}}}{\pi  {\left(1-v^{2 \sqrt{\frac{2}{3}}}\right)^{1/4}}}~.
\end{array}
\eea
The longitudinal Langevin coefficients for oblate and prolate backgrounds are
\bea
\begin{array}{l}
\k_{L,obl}^{\parallel}=\frac{{2^{1/4}} v^{1-\frac{1}{2 \sqrt{2}}} \left(-\left(\sqrt{2}-1\right) v^{2 \sqrt{2}}+\sqrt{2}+1\right)^{3/2}}{\pi ^2 \left(1-v^{2 \sqrt{2}}\right)^{5/4}}\,, \quad
\k_{L, pro}^{\parallel}=\frac{\sqrt[4]{\frac{2}{3}} v^{-\frac{1}{2}-\frac{1}{2 \sqrt{6}}} \left(v^{2 \sqrt{\frac{2}{3}}}+1\right)^{3/2}}{\pi ^2 \left(1-v^{2 \sqrt{\frac{2}{3}}}\right)^{5/4}}
\,,\\\\
\k_{L,obl}^{\perp}=\frac{v^{-\frac{1}{3 \sqrt{2}}} \left(-\left(\sqrt{2}-2\right) v^{\frac{4 \sqrt{2}}{3}}+\sqrt{2}+2\right)^{3/2}}{2^{3/4} \sqrt{3} \pi ^2 \left(1-v^{\frac{4 \sqrt{2}}{3}}\right)^{5/4}}\,,\qquad
\k_{L,pro}^{\perp}=\frac{v^{1-\frac{1}{\sqrt{6}}} \left(-\left(\sqrt{6}-2\right) v^{4 \sqrt{\frac{2}{3}}}+\sqrt{6}+2\right)^{3/2}}{2^{3/4} \sqrt[4]{3} \pi ^2 \left(1-v^{4 \sqrt{\frac{2}{3}}}\right)^{5/4}} \,,
\end{array}
\eea
and the transverse Langevin coefficients  are
\bea
\begin{array}{l}
\k_{T,obl}^{\perp,(\parallel)}=\frac{\sqrt{3} v^{\frac{1}{6} \left(4-\sqrt{2}\right)} {\left(\sqrt{2}+2-\left(\sqrt{2}-2\right) v^{\frac{4 \sqrt{2}}{3}}\right)^{1/2}}}{\pi ^2 {\left(2-2 v^{\frac{4 \sqrt{2}}{3}}\right)^{1/4}}}  \,,\quad \k_{T,pro}^{\perp,(\parallel)}=\frac{{\left(\frac{3}{2}\right)^{1/4}} v^{-1-\frac{1}{\sqrt{6}}} {\left(\sqrt{6}+2-\left(\sqrt{6}-2\right) v^{4 \sqrt{\frac{2}{3}}}\right)^{1/2}}}{\pi ^2 {\left(1-v^{4 \sqrt{\frac{2}{3}}}\right)^{1/4}}}  \,,\\\\
\k_{T,obl}^{\perp,(\perp)}=\frac{\sqrt{3} v^{-\frac{1}{3 \sqrt{2}}} {\left(\sqrt{2}+2-\left(\sqrt{2}-2\right) v^{\frac{4 \sqrt{2}}{3}}\right)^{1/2}}}{\pi ^2 {\left(2-2 v^{\frac{4 \sqrt{2}}{3}}\right)^{1/4}}}  \,, \quad \k_{T,pro}^{\perp,(\perp)}=\frac{\sqrt[4]{\frac{3}{2}} v^{1-\frac{1}{\sqrt{6}}} {\left(\sqrt{6}+2-\left(\sqrt{6}-2\right) v^{4 \sqrt{\frac{2}{3}}}\right)^{1/2}}}{\pi ^2 {\left(1-v^{4 \sqrt{\frac{2}{3}}}\right)^{1/4}}}\,,  \\\\
\k_{T,obl}^{\parallel,(\perp)}=\frac{2^{3/4} v^{-\frac{1}{2 \sqrt{2}}} {\left(\sqrt{2}+1-\left(\sqrt{2}-1\right) v^{2 \sqrt{2}}\right)^{1/2}}}{\pi ^2 {\left(1-v^{2 \sqrt{2}}\right)^{1/4}}}  \,,\quad \k_{T,pro}^{\parallel,(\perp)}=\frac{2^{3/4} \,{3^{1/4}} v^{\frac12-\frac{1}{2 \sqrt{6}}} { \left(v^{2 \sqrt{\frac{2}{3}}}+1\right)^{1/2}}}{\pi ^2 {\left(1-v^{2 \sqrt{\frac{2}{3}}}\right)^{1/4}}}  \,.
\end{array}
\eea

We also summarize the results for isotropic background in bottom-up theory with fixed $A=1$ and $B=0$. These read
\bea
\begin{array}{c c}
u_{0,iso}=\frac{\left(1-v\right)^{1/4}}{\left(1+v\right)^{1/4}}~,~~~~ & T_{ws,iso}=\frac{\sqrt{2} \left(1-v^2\right)^{1/4}}{\pi }~,\\\\
\kappa_{_{L,iso}}=\frac{2 \sqrt{2}}{\pi ^2 \left(1-v^2\right)^{5/4}}~,~~~~ &
\kappa_{_{T,iso}}=\frac{2 \sqrt{2}}{\pi ^2 \left(1-v^2\right)^{1/4}}~.
\end{array}\label{iso-jw}
\eea

\bibliographystyle{JHEP}
\bibliography{botany}

\end{document}